# Verification and Validation of the Stakeholder Tool for Assessing Radioactive Transportation (START) – 22323

Caitlin Condon *, Philip Jensen *, Patrick Royer *, Harish Reddy Gadey *, Mark Abkowitz **,
Robert Claypool ***, Steven Maheras *, Matt Feldman *
\* Pacific Northwest National Laboratory
\*\* Vanderbilt University
\*\*\* Kanini Software Solutions

ABSTRACT

The U.S. Department of Energy (DOE) Office of Integrated Waste Management is planning for the eventual transportation, storage, and disposal of spent nuclear fuel (SNF) and high-level radioactive waste (HLW) from nuclear power plant and DOE sites. The Stakeholder Tool for Assessing Radioactive Transportation (START) is a web-based, geospatial decision-support tool developed for evaluating routing options and other aspects of transporting SNF and HLW, covering rail, truck, barge, and intermodal infrastructure and operations in the continental United States. The verification and validation (V&V) process is intended to independently assess START to provide confidence in the ability of START to accurately provide intended results. The V&V process checks the START tool using a variety of methods, ranging from independent hand calculations to comparison of START performance and results to those of other codes.

The V&V activity was conducted independently from the START development team with opportunities to provide feedback and collaborate throughout the process. The V&V analyzed attributes of transportation routes produced by START, including route distance and both population and population density captured within buffer zones around routes. Population in the buffer zone, population density in the buffer zone, and route distance were all identified as crucial outputs of the START code and were subject to V&V tasks. Some of the improvements identified through the V&V process were standardizing the underlying population data in START, changing the projection of the population raster data, and changes to the methodology used for population density to improve its applicability for expected users. This collaboration also led to suggested improvements to some of the underlying shape file segments within START.

A formal series of test routes went through the V&V process from START version 3.2.1 following the implementation of the recommended improvements to START. The V&V of the buffer zone population and distance reported in START were achieved by utilizing a custom workflow developed in QGIS. This platform was chosen because it is fully independent of START. The buffer zone population and route distances in START showed excellent agreement to independent V&V test results. Over 200 route test cases were run; in all cases, the percent difference in the population within the buffer zone was less than +/- 5%, and the majority of cases were below +/- 1%. These cases were also duplicated in ArcMap, which is Environmental Systems Research Institute's (ESRI) GIS desktop application. The ArcMap results also show good agreement; however, they cannot be considered independent of START because START also uses ESRI software.

The V&V tasks are currently addressing population density and estimated radiation dose in START as well as implementing further automation of the V&V process to allow larger suites of test routes to be run more easily and quickly by less experienced GIS users. These improvements to the V&V process will make it a valuable tool for the development team and the stakeholders so that each new iteration of START can easily be checked for quality and consistency. This paper and presentation will describe the V&V process, findings and implications, and ongoing V&V activities. PNNL-SA-169303.

**INTRODUCTION[1]**

The United States DOE Office of Integrated Waste Management is planning for the eventual transportation, storage, and disposal of spent nuclear fuel (SNF) and high-level radioactive waste (HLW) generated at nuclear power plant and DOE sites across the United States. To aid in this effort, the DOE developed the Stakeholder Tool for Assessing Radioactive Transportation (START), a web-based decision-support tool. This supports efforts to evaluate transportation route options and other analysis and planning activities associated with the transportation of radioactive materials between sites [1].

As a V&V process can be used to assess confidence in code outputs, an independent V&V effort for START was therefore initiated at PNNL. Several methods were employed ranging from independent calculations to comparison of START performance with other codes as part of the V&V effort. The V&V process details the methods employed to independently check START outputs for route buffer zone population, buffer zone population density, and distance against independent assessments in ArcMAP [2] and QGIS [3]. V&V of software codes is a crucial task for ensuring software quality. The following paper summarizes recent work to benchmark the START code [4,5].

**DISCUSSION**

**Agile Software Development**

The collaborative efforts between the PNNL V&V team and the START development team has led to an increased understanding of the challenges START faces on the GIS front and the identification of potential scope for improvements to better support START user and stakeholder needs. The communication methodology employed between the START V&V team and the development team is a form of agile software development. Testing done by the V&V team leads to feedback provided to the software developers which leads to improvements in the software, which leads to a new round of V&V tests. The concept of agile software development is depicted in Fig. 1.

---

[1] This is a technical paper that does not take into account contractual limitations or obligations under the Standard Contract for Disposal of Spent Nuclear Fuel and/or High-Level Radioactive Waste (Standard Contract) (10 CFR Part 961). For example, under the provisions of the Standard Contract, spent nuclear fuel in multi-assembly canisters is not an acceptable waste form, absent a mutually agreed to contract amendment.
To the extent discussions or recommendations in this paper conflict with the provisions of the Standard Contract, the Standard Contract governs the obligations of the parties, and this paper in no manner supersedes, overrides, or amends the Standard Contract.
This paper reflects technical work which could support future decision making by the Department of Energy (DOE or Department). No inferences should be drawn from this paper regarding future actions by DOE, which are limited both by the terms of the Standard Contract and Congressional appropriations for the Department to fulfil its obligations under the Nuclear Waste Policy Act including licensing and construction of a spent nuclear fuel repository.

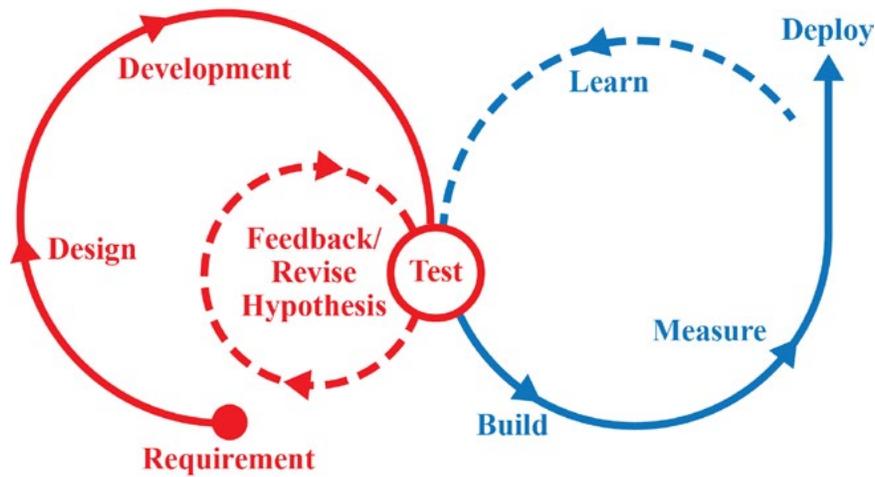

**Fig. 1. The concept of agile software development**

Over the course of Fiscal Year 2020, the PNNL V&V team conducted multiple tests to check the population in the buffer zones (800-meters and 2500-meters) as well as report on total route distances. Any inconsistencies found were shared with the START development team to be remedied. After each update in START, the PNNL V&V team performed an independent series of tests until no inconsistencies were detected. Results presented in this work refer to the final version of START available in Fiscal Year 2020 (START 3.2.1), but V&V testing is still ongoing for subsequent versions of START.

**Initial V&V Testing**

In the first rounds of testing, the V&V team identified inconsistencies in the total route population compared to the V&V tests. It was observed that when the START total route function was used, the population estimates were consistently lower than those derived from the V&V efforts. START also presents individual segment level information to the users. When a similar exercise was conducted at the segment level, it was observed that the results from START and the V&V team were similar. Further investigation led to the conclusion that the segment level population data was being extracted from the LandScan [6] raster population data while the START total route population was being extracted from the population census data. Routes are a collection of individual segments. The buffer zone populations for individual segments are pre-defined while the full route population data is calculated after the user defined route is developed. The population for a given route is determined by the intersection of the polygon and the centroid points of the underlying population data. The route polygon was less likely to capture the centroid of a census block in the census data compared to the LandScan raster data, thereby leading to different population estimates. This discrepancy was rectified by the START development team and all population estimates were updated to be obtained from LandScan raster data rather than census data.

Following this update, the V&V team noticed there were still some inconsistencies between the START output for population in the buffer zone and what was estimated in QGIS and ArcMAP. The V&V team worked closely with the development team to investigate the discrepancy and determined that the difference in estimates was caused by the map projection used for full routes in the START tool. The START development team resolved this projection discrepancy and V&V testing recommenced.

Following this update, the V&V team found close alignment between START outputs and the independent assessments for population within the buffer zone and route distances with a few exceptions. The V&V efforts identified that there were differences in the population results when working from either a downloaded KML file or a downloaded Shape file for the same START test route. The V&V efforts found that V&V test results working from the KML file downloads from START had presented lower percent differences than the corresponding V&V test results from Shape files. Upon close inspection of the Shape file routes that were presenting abnormal population results through the V&V test process, it was discovered that some of the START Shape files presented minor abnormalities, which affected the independent V&V efforts for population and route distance independently of START. An example of a route aberration is shown in Fig. 2. The START development team is currently working on addressing this issue in the Shape file downloads. For all V&V efforts for START version 3.2.1, the KML files were used.

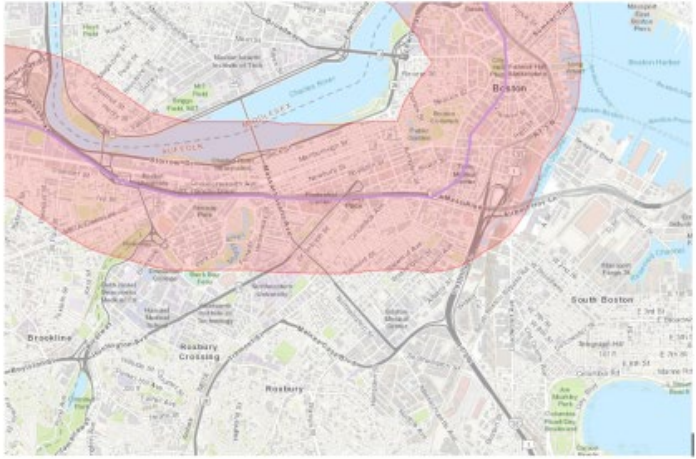
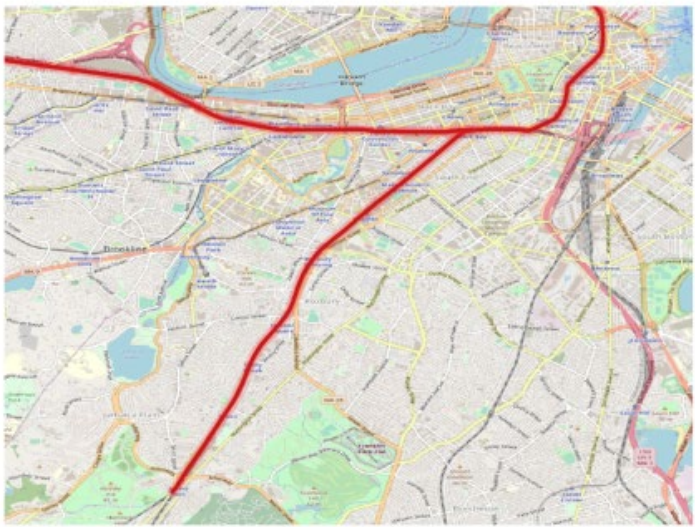
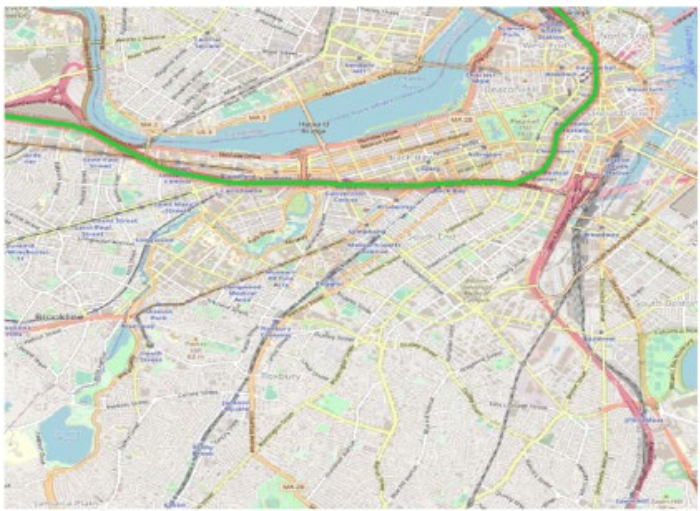

**Fig.2. Example of START Route Shape File Aberration**

**Test Route Selection**

Following the initial testing, an official set of test routes was developed for this V&V effort of START version 3.2.1 as well as V&V efforts for future versions of START. The origin point of each test route was a shutdown nuclear power plant site, nuclear facility, or DOE facility. The destination of all the routes was the geographic continental center of the United States. Each of these routes was tagged with a unique testing identification number. Each origin-destination pair was developed into two routes for each of the two buffer zone options: one at 800-meters and the second at 2500-meters. All the routes were run using the minimum distance criteria. As for the mode of transportation, the following priority was used for the origin destination pairs depending on what mode was available for each: (1) Rail only, (2) Heavy haul truck to rail, (3) Heavy haul truck, and (4) Barge to rail. The test series includes over 300 unique test routes. Not all origin-destination pairs in START successfully generated a test route; 19 origin destination pairs failed to generate a valid route through START using any mode of transport.

**Route Buffer Zone Population Calculations**

The buffer zone population V&V was performed using two independent GIS tools: QGIS and ArcMap. QGIS was used since it is a completely independent GIS tool and does not utilize any ESRI software. ArcMap was used to check the results in a separate GIS program but with the same underlying ESRI software as START. Workflows were developed and Python scripts were written to partially automate the process for the V&V efforts. Population data was extracted for all the routes for both 800- and 2500-meter buffer zone distances from the LandScan raster population data sets for both day and at night. The average population for a route was calculated using eq (1).

$$Average\ Population = \frac{Day + (2*Night)}{3} \qquad \text{(Eq. 1)}$$

The average population extracted from QGIS and ArcMap were then compared to the START results as part of the V&V process. Strong agreement was observed between the START results and the ArcMap data; this result was expected as START uses ESRI features for calculations and ArcMAP is an ESRI product. On comparing the difference between the START results and QGIS, it was observed that the differences were within 5% with most cases falling within 1% of each other. It is worth noting that the QGIS results are completely independent of START. The full results and exact calculations can be found in the START V&V reports [4,5].

**Route Distance Calculations**

The route distance was verified and validated using the QGIS software. The route must be reprojected from the native ESPG:4326 projection into ESPG:3857 or web Mercator in order to calculate the route distance (the population calculations are all conducted in the native raster projection of 4326). This can be accomplished in QGIS by applying the projection tool to convert the polyline route from the native projection to a user-defined projection. The user can then go into the attribute table of the reprojected route and use the field calculator tool to obtain the route distance.

The route distance V&V was conducted in QGIS because it is fully independent of START. In addition, START utilizes the ArcMap Network Analyst tool for this calculation. The route distance calculations available to the user in START are not easily replicated in the user interface of ArcMap without the Network Analyst tool. The user would need to develop a python work-around to replicate this calculation in the same projection that START utilizes. For all routes in the test route series the percent difference between the route distance reported in START and the route distance estimated in QGIS were well below +/- 5% with most cases being well below +/- 1%.

**Route Buffer Zone Population Density Assessment**

The V&V team also performed work testing the population density for a full route reported in START. The V&V team discovered inconsistencies with the START-reported population density and the V&V estimated population density for a corresponding route. The START development team and the V&V team met to determine where the inconsistencies in the population density were originating. It was determined that the population density reported by START is based on a weighted combination of the individual route segment population densities. The START development team and the V&V team worked together to determine the most accurate method for calculating population density and concluded that the population density would be best calculated across the total route rather than based on underlying route segments. This avoids the possibility of mis-weighting population densities based on buffer zone overlaps for individual route segments. The START development team and the V&V team believe that this coding change for population density calculations will lead to increased user options, such as obtaining population density per State or specified region along the route. This coding change is under development by the START development team and will be checked by the V&V team after it is finalized.

**CONCLUSIONS**

The V&V efforts for START version 3.2.1 found good agreement between START results and independent assessments for full route buffer zone population and route distance. The V&V utilized QGIS as a fully independent GIS tool and ArcMap, which is not independent from START but provides confirmation that START outputs can be replicated in other ESRI products. In all test cases, the route buffer zone population and route distance showed good agreement between START and the V&V test cases. The percent difference was less than +/- 5% in all cases, with the majority of cases less than +/- 1%.

V&V efforts are ongoing for updates in START as well as new features that have become available in START version 3.2.2. The START V&V team as well as the START development team will continue collaborative efforts to identify areas of potential improvement within START. The ongoing V&V tasks include addressing population density and estimated radiation dose in START as well as implementing further automation of the V&V process to allow larger suites of test routes to be run more easily and quickly by less experienced GIS users. These improvements to the V&V process will make it a valuable tool for the development team and the stakeholders so that each new iteration of START can easily be checked for quality and consistency.

## ACKNOWLEDGEMENTS

Pacific Northwest National Laboratory is operated by Battelle Memorial Institute for the US Department of Energy under Contract No. DE-AC05-76RL01830. This work was supported by the US Department of Energy Office of Integrated Waste Management.